\documentclass[amssymb,twocolumn,aps,eqsecnum]{revtex4}
\usepackage{amsmath}
\usepackage{enumerate}
\begin{document}
\title{Electromagnetic Energy, Momentum, and Angular Momentum in an
Inhomogeneous Linear Dielectric}
\author{Michael E. Crenshaw and Thomas B. Bahder}
\affiliation{US Army Aviation and Missile Research, Development, and Engineering Center, Redstone Arsenal, AL 35898, USA}
\date{\today}
\begin{abstract}
In a previous work, Optics Communications 284 (2011) 2460--2465,
we considered a dielectric medium with an anti-reflection coating and a
spatially uniform index of refraction illuminated at normal incidence
by a quasimonochromatic field.
Using the continuity equations for the electromagnetic energy density
and the Gordon momentum density, we constructed a traceless, symmetric
energy--momentum tensor for the closed system.
In this work, we relax the condition of a uniform index of refraction
and consider a dielectric medium with a spatially varying index of
refraction that is independent of time, which essentially represents a
mechanically rigid dielectric medium due to external constraints.
Using continuity equations for energy density and for Gordon momentum
density, we construct a symmetric energy--momentum matrix, whose
four-divergence is equal to a generalized Helmholtz force density 
four-vector.
Assuming that the energy-momentum matrix has tensor transformation
properties under a symmetry group of space-time coordinate
transformations, we derive the global conservation laws for the total
energy, momentum, and angular momentum.
\end{abstract}
\maketitle
\vskip 3.0cm
\par
\section{Introduction}
\par
Starting with the hydrodynamic continuity equation, Umov \cite{BIUmov}
obtained an expression for energy continuity in a continuous spatial
flow of an electromagnetic field in
1874 \cite{BIHeaviside,BIPoynting,BIMicrophysics}.
A decade later, Poynting \cite{BIPoynting} derived a similar energy
continuity equation as a general theorem of the macroscopic Maxwell
equations.
Poynting's theorem is generally preferred because it can be derived
directly from the macroscopic Maxwell equations
\begin{equation}
\nabla\times {\bf B}-\frac{n^2}{c}\frac{\partial {\bf E}}{\partial t}=0
\label{EQq1.01}
\end{equation}
\begin{equation}
\nabla\times {\bf E}+\frac{1}{c}\frac{\partial{\bf B}}{\partial t} =0
\label{EQq1.02}
\end{equation}
\begin{equation}
\nabla \cdot ( {n^2 \bf{E}} )= 0
\label{EQq1.03}
\end{equation}
\begin{equation}
\nabla \cdot {\bf{B}} = 0
\label{EQq1.04}
\end{equation}
of classical continuum electrodynamics.
Here, the Maxwell equations are written in Heaviside--Lorentz units
for a nonmagnetic linear medium in the absence of free charges and
currents.
We have assumed that the index of refraction, $n=n({\bf r})$,
depends on position, occupies a finite region of 3-dimensional
space, and is independent of time.
Poynting's theorem can be derived by subtracting the scalar product
of the Maxwell--Amp\`ere law,
Eq.\ (\ref{EQq1.01}), with ${\bf E}$ from the scalar product of
Faraday's law, Eq.\ (\ref{EQq1.02}), with ${\bf B}$ to obtain
$$
\frac{n^2}{c}\frac{\partial {\bf E}}{\partial t} \cdot {\bf E}+
\frac{1}{c}\frac{\partial {\bf B}}{\partial t} \cdot {\bf B}
$$
\begin{equation}
+(\nabla\times {\bf E})\cdot {\bf B}
-(\nabla\times {\bf B})\cdot {\bf E}
=0.
\label{EQq1.05}
\end{equation}
Upon application of a vector identity and the definition of the energy
density, 
\begin{equation}
\rho_e=\frac{1}{2}(n^2 {\bf E}^2+{\bf B}^2),
\label{EQq1.06}
\end{equation}
the preceding equation becomes Poynting's
theorem
\begin{equation}
\frac{\partial \rho_e}{\partial t}+\nabla\cdot c({\bf E}\times{\bf B})=0
\label{EQq1.07}
\end{equation}
and defines Poynting's energy flux
vector ${\bf S}_P=c({\bf E}\times{\bf B})$.
\par
In a previous work \cite{BICB}, we considered a dielectric medium with
an index of refraction that was time-independent, spatially uniform,
and covered with a thin gradient-index antireflection coating.
For this case, we showed that the energy
\begin{equation}
U=\int_{\sigma} \frac{1}{2}\left ( n^2{\bf E}^2+{\bf B}^2 \right ) dv
\label{EQq1.08}
\end{equation}
and the Gordon \cite{BIGord} momentum
\begin{equation}
{\bf G}_G=\int_{\sigma} \frac{1}{c} (n{\bf E} \times {\bf B}) dv
\label{EQq1.09}
\end{equation}
are the conserved electromagnetic quantities with integration performed
over all space $\sigma$.
We constructed the corresponding traceless, symmetric energy--momentum
tensor, whose four--divergence provided the continuity equations for the
energy and momentum densities of the closed system.
This result incorporated a condition of $\nabla n / n\ll 1/\lambda$
corresponding to unimpeded flow in the absence of external forces.
\par
In the current work, we consider the case of an inhomogeneous dielectric
medium in which the condition $\nabla n/n\ll 1/\lambda$ no
longer holds.
The assumption that the index of refraction is independent of time 
necessarily implies that the dielectric medium is mechanically
rigid~\cite{RigidFootnote} and that no momentum is transferred to the
dielectric medium. 
In other words, at each spatial point, there is {\it effectively} an
external field that acts as a constraint that holds the dielectric in
place. 
Therefore, the system that we are considering is not a closed system
and the continuity equations and global conservation equations will
reflect this feature.
The external field that acts as a constraint is not a field that we
impose, instead it arises as a result of the way in which we arrange
terms in the continuity equations.
We construct the energy and momentum continuity equations and find that
the spatial gradient of the refractive index appears in the continuity
equations in a generalized Helmholtz force density.
This generalized Helmholtz force density is a spatial and time-varying
field that provides the constraint for the dielectric medium to be
mechanically rigid.
We then write the continuity equation
as a four divergence of a symmetric energy--momentum matrix. 
Assuming that the energy-momentum matrix has tensor transformation
properties under a symmetry group of space-time coordinate
transformations, we apply the four-dimensional divergence theorem to
derive global conservation laws for the total
energy, momentum, and angular momentum.
\par
\section{Energy and Momentum Continuity Equations}
\par
The energy--momentum tensor is a concise way to represent the local 
continuity of energy and momentum of a field.
While simple in concept, the form of the energy--momentum tensor for the
electromagnetic field in a dielectric has been at the center of the
century-long Abraham-Minkowski controversy \cite{BIPfeifer}. 
The tensor that was proposed by Minkowski \cite{BIMin} in 1908 is not
diagonally symmetric, a fact that is adverse to conservation of angular 
momentum \cite{BILL}.
Abraham \cite{BIAbr} subsequently proposed a symmetric tensor at the
expense of a phenomenological force.
The disagreeable properties of the Minkowski and Abraham energy-momentum
tensors are manifestations of underlying conservation issues:
neither the Minkowski momentum nor the Abraham momentum is conserved.
In this section, we derive the energy and momentum continuity equations
for an inhomogeneous dielectric and construct the corresponding tensor
continuity equation and energy--momentum tensor.
In the following section, we will obtain the globally conserved
quantities---the total energy, the total momentum, and the total
angular momentum system, from the energy momentum tensor.
\par
Starting from the Maxwell Eqs.(\ref{EQq1.01})--(\ref{EQq1.04}), we
use the continuity of the electromagnetic energy density and the
Gordon momentum to obtain a symmetric stress-energy tensor.
We begin with the temporal derivatives of the energy density
and the Gordon momentum density,
\begin{equation}
\frac{n}{c}\frac{\partial}{\partial t}
\frac{1}{2}\left (n^2{\bf E}^2+{\bf B}^2\right )=
\frac{n}{c}\frac{\partial (n {\bf E}) }{\partial t}\cdot n{\bf E} +
\frac{n}{c}\frac{\partial {\bf B} }{\partial t}\cdot {\bf B} 
\label{EQq2.01}
\end{equation}
\begin{equation}
\frac{n}{c}\frac{\partial}{\partial t}(n{\bf E}\times{\bf B})=
\frac{n}{c}\frac{\partial (n {\bf E}) }{\partial t}\times{\bf B} +
n{\bf E}\times\frac{n}{c}\frac{\partial {\bf B} }{\partial t}.
\label{EQq2.02}
\end{equation}
The Gordon momentum in Eq.~(\ref{EQq2.02}) has been scaled by $c$
so that the equations are in the same units.
Next, we apply the vector identity 
\begin{equation}
\nabla\times(\psi{\bf a})
=\nabla\psi\times{\bf a}+\psi\nabla\times{\bf a}
\label{EQq2.03}
\end{equation}
to Faraday's law and write the macroscopic Maxwell equations,
Eqs.~(\ref{EQq1.01}) and (\ref{EQq1.02}), as
\begin{equation}
\frac{n}{c}\frac{\partial (n{\bf E})}{\partial t}=\nabla\times {\bf B}
\label{EQq2.04}
\end{equation}
\begin{equation}
\frac{n}{c}\frac{\partial{\bf B}}{\partial t}=
-\nabla\times (n{\bf E})+\frac{\nabla n}{n}\times n{\bf E}.
\label{EQq2.05}
\end{equation}
The variant form of Maxwell's equations, Eqs.~(\ref{EQq2.04})
and (\ref{EQq2.05}), are mathematically equivalent to the original
versions, Eqs.~(\ref{EQq1.01}) and (\ref{EQq1.02}), respectively.
Substituting the Maxwell equations, Eqs.~(\ref{EQq2.04}) and
(\ref{EQq2.05}), into Eqs.~(\ref{EQq2.01}) and (\ref{EQq2.02}), we
produce the energy continuity equation
\begin{equation}
\frac{n}{c}\frac{\partial\rho_e}{\partial t}
+\nabla\cdot (n{\bf E}\times{\bf B})
=\frac{\nabla n}{n}\cdot (n{\bf E}\times{\bf B})
\label{EQq2.06}
\end{equation}
and the (Gordon) momentum continuity equation
$$
\frac{n}{c}\frac{\partial}{\partial t}(n{\bf E}\times{\bf B})
+\nabla\cdot{\bf W}
+n{\bf E}\left (\nabla\cdot n{\bf E} \right )
=
$$
\begin{equation}
\left ( n{\bf E}\times\frac{\nabla n}{n}\right )\times n{\bf E}.
\label{EQq2.07}
\end{equation}
Here, we have used Eq.~(\ref{EQq1.03}) and the definition of the
Maxwell stress tensor \cite{BIJackson,BILL}
\begin{equation}
W_{ij}
=\left (-nE_{i}nE_{j}-B_{i}B_{j}+
\frac{1}{2}\left ( n{\bf E}\cdot n{\bf E}+
{\bf B}\cdot{\bf B}\right )\delta_{ij}\right ).
\label{EQq2.08}
\end{equation}
\par
We can write the energy continuity equation, Eq.~(\ref{EQq2.06}), and
momentum continuity equation, Eq.~(\ref{EQq2.07}), as the matrix
differential equation
\begin{equation}
\bar\partial_{\beta}T^{\alpha\beta}=f^{\alpha}
\label{EQq2.09}
\end{equation}
with summation over repeated indices.
The quantities that appear in Eq.~(\ref{EQq2.09}) are a
four-divergence operator
\begin{equation}
\bar\partial_{\alpha}=\left (
\frac{n}{c}\frac{\partial}{\partial t},
\partial_x,\partial_y,\partial_z
\right ),
\label{EQq2.10}
\end{equation}
the array
\begin{equation}
T^{\alpha\beta}= \left [
\begin{matrix}
\rho_e
&c{g}_{{\rm G}_1}
&c{g}_{{\rm G}_2}
&c{g}_{{\rm G}_3}
\cr
c{g}_{{\rm G}_1}
&W_{11}
&W_{12}
&W_{13}
\cr
c{g}_{{\rm G}_2}
&W_{21}
&W_{22}
&W_{23}
\cr
c{g}_{{\rm G}_3}
&W_{31}
&W_{32}
&W_{33}
\cr
\end{matrix}
\right ] ,
\label{EQq2.11}
\end{equation}
and a generalized force density four-vector
\begin{equation}
f^{\alpha}= \left (
\nabla n \cdot ( {\bf E}\times{\bf B}), -{\bf f}_H
\right ),
\label{EQq2.12}
\end{equation}
where
\begin{equation}
{\bf f}_H=-nE^2\nabla n +2n{\bf E}\left ({\bf E}\times\nabla n \right )
+ n^2 {\bf E} ( \nabla \cdot {\bf E} ).
\label{EQq2.13}
\end{equation}
Using Eq.~(\ref{EQq1.03}), the last two terms on the right side of
Eq.~(\ref{EQq2.13}) cancel, so that
\begin{equation}
{\bf f}_H = - n {E^2}\,\nabla {n} 
\label{EQq2.14}
\end{equation}
in the absence of free charges.
\par
The array in Eq.~(\ref{EQq2.11}) appears to have the properties of 
an energy--momentum tensor.
By construction, the operator defined in Eq.\ (\ref{EQq2.10}) applied
to the rows of the array in Eq.\ (\ref{EQq2.11}) generates continuity
equations for the electromagnetic energy and the momentum.
The same operation applied to the columns
\begin{equation}
\bar\partial_{\alpha} T^{\alpha\beta}=f^{\beta},
\label{EQq2.15}
\end{equation}
generates the same continuity equations by symmetry
\begin{equation}
T^{\alpha\beta}=T^{\beta\alpha}.
\label{EQq2.16}
\end{equation}
The array has a vanishing trace
\begin{equation}
T^{\alpha}_{\alpha}= 0
\label{EQq2.17}
\end{equation}
corresponding to massless particles \cite{BIJackson,BILL}.
\par
The new feature of this result is the
appearance of the four-vector $f^{\alpha}$ in the continuity
equation (\ref{EQq2.09}).
The appearance of the four-vector $f^{\alpha}$ in the divergence of the
stress energy tensor in Eq.~(\ref{EQq2.09}) is a result of the fact that
the system is not closed.
When the gradient of the index of refraction is non-zero, there is a
back-action on the field altering its spatial properties, so the field
does not experience ``unimpeded flow''.
It is possible to define an effective stress-energy tensor that takes
into account the back-action on the field by introducing
\begin{equation}
\partial_\alpha t^{\alpha \beta} = f^\beta .
\label{EQq2.18}
\end{equation}
The continuity of energy and momentum can then be expressed by 
\begin{equation}
\partial_\alpha (T^{\alpha \beta} + t^{\alpha \beta}) = 0
\label{EQq2.19}
\end{equation}
where the tensor $t^{\alpha \beta}$ contains the influence of the
inhomogeneity of the material and is zero for homogeneous dielectrics.
\par
\section{Global Conservation Equations}
\par
If we assume that $T^{\alpha\beta}$ transforms as a tensor under some
symmetry group of space-time coordinate transformations, then we can
apply the four-dimensional divergence theorem to obtain global
conservation equations.
Integrating Eq.~(\ref{EQq2.09}) over a four-volume,
$d\Omega = c d {\bar t} \; d^3x \equiv c d {\bar t} \; dv $
(where ${\bar t} = t/n$), between hypersurfaces~\cite{BILL} of
constant time at $\bar t_1=t_1/n$ and $\bar t_2=t_2/n$, we have
\begin{equation}
\int_{\bar t_1}^{\bar t_2}
\bar\partial_{\beta} T^{\alpha 0} \; cd\bar t \; d v
=\int_{\bar t_1}^{\bar t_2}
 f^\alpha \; cd \bar t \; d v,
\label{EQq3.01}
\end{equation}
where the integrals over $d v =d^3x$ are 3-dimensional volume integrals
over the volume containing the field.
Applying the four-divergence theorem results in
\begin{equation}
\int \left( T^{\alpha 0} (t_2) - T^{\alpha 0} (t_1) \right) \, d v
= c\int \; \frac{1}{n} f^\alpha d t \; d v.
\label{EQq3.02}
\end{equation}
When $\alpha = 1,2,3$,
Eq.~(\ref{EQq3.02}) reduces to 
\begin{equation}
{\bf G}(t_2) - {\bf G} (t_1) = \int \, E^2 \, \nabla n \, d t \; d v
\label{EQq3.03}
\end{equation}
where ${\bf G}(t) $ is the Gordon momentum at time $t$, and the time
integration is between $t=t_1$ and $t=t_2$.
Equation~(\ref{EQq3.03}) shows that the
Gordon momentum of the field is not constant, i.e., there is a source
or sink of momentum provided by the 
external constraint field ${\bf f}_H$.
For the case $\alpha =0$, Eq.~(\ref{EQq3.02}) gives
\begin{equation}
\int \left ( \rho_e(t_2) - \rho_e(t_1) \right ) d v =
\int \, \frac{\nabla n}{n} \cdot ( {\bf E}\times{\bf B}) \, d t \; d v
\label{EQq3.04}
\end{equation}
where $\rho_e$ is given by Eq.~(\ref{EQq1.06}).
The left side of Eq.~(\ref{EQq3.04}) is the difference in total
energy at two different times, $U(t_2)-U(t_1)$.
We see that a non-zero spatial gradient in the index means that the
field can gain or loose energy.
The results given in Eqs.~(\ref{EQq3.03}) and (\ref{EQq3.04})
assume that the matrix $T^{\alpha\beta}$ in Eq.~(\ref{EQq2.09})
transforms as a tensor under some symmetry group of space-time
coordinate transformations.
Note that when $\nabla n=0$, Eqs.~(\ref{EQq3.03}) and (\ref{EQq3.04})
show that the Gordon momentum ${\bf G}(t)$ and the energy $U(t)$ are
constants. 
\par
In recent years, there is intense interest in angular momentum
carried by the electromagnetic field, see the recent review and
references cited therein~\cite{TwistedPhotons,BIJHL}. 
However, the angular momentum carried by the electromagnetic field in
a dielectric environment is no less unsettled than the linear momentum
case \cite{BIang}.
We can define the four-tensor of angular momentum density
in terms of our energy--momentum tensor as~\cite{BILL}
\begin{equation}
m^{\alpha \beta \gamma} = \frac{1}{c} \left( x^\alpha \,
T^{\gamma \beta} - x^\beta \, T^{\gamma \alpha} \right) =
- m^{\beta \alpha \gamma}.
\label{EQq3.05}
\end{equation}
Continuity of angular momentum is given by 
\begin{equation}
\bar \partial_\gamma m^{\alpha \beta \gamma} =
\frac{1}{c} \left( x^\alpha \, f^\beta - x^\beta \, f^\alpha \right).
\label{EQq3.06}
\end{equation}
The divergence of $m^{\alpha \beta \gamma}$ is not zero, thereby
indicating that there is a source or sink of angular momentum density, 
due to the gradient of the index of refraction.
\par
Once again, we assume that there exists a symmetry group of space-time
coordinate transformations, so that $m^{\alpha \beta \gamma}$ is a
tensor. 
As above, we can then use the four-divergence theorem to obtain
\begin{equation}
\int\left ( m^{\alpha \beta 0}(t_2)-m^{\alpha \beta 0} (t_1) \right )
\, dv = \int \frac{1}{n} \left( x^\alpha f^\beta -
x^\beta f^\alpha \right) dt \, dv
\label{EQq3.07}
\end{equation}
where the time integral is between the two times $t_1$ and $t_2$ and
where the volume integral $dv $ is over the portion of
three-dimensional space containing the field and includes the region
where $n(\bf r) > 1$.
When $\alpha$ and $\beta$ take values \mbox{$i,j=1,2,3$}, we have 
\begin{equation}
 m^{i j 0} (t) = x^i f^j - x^j f^i
\label{EQq3.08}
\end{equation}
where $f^i$ are the components defined in Eq.~(\ref{EQq2.14}).
Equation~(\ref{EQq3.07}) gives the change in total 
angular momentum of the field, $\Delta {\bf M}$, between time $t_1$
and $t_2$, and can be written as
\begin{equation}
\Delta {\bf M} = \int \left ( {\bf m} (t_2) - {\bf m} (t_1) \right )
\, dv = \int E^2 \left( {\bf r} \times \nabla n \right) dt \, dv
\label{EQq3.09}
\end{equation}
where the ${\bf m} (t)$ is the angular momentum density at
position ${\bf r}$ at time $t$, defined by
\begin{equation}
 {\bf m} (t) = {\bf r} \times {\bf g_G}(t)
\label{EQq3.10}
\end{equation}
and is defined in terms of the Gordon momentum density
\begin{equation}
{\bf g_G}(t)= n {\bf E} \times {\bf B}/c.
\label{EQq3.11}
\end{equation}
Equation~(\ref{EQq3.09}) shows that when ${\bf r} \times \nabla n$ 
is non-zero, then the total angular momentum of the field
(as expressed through the angular momentum density) can change. 
Once again, we remind the reader that we have assumed the index of
refraction as isotropic, constant in time, but varying in position.
Equation~(\ref{EQq3.09}) essentially shows that a
particular spatial distribution of 
${\bf r} \times \nabla n$ can lead to a back-action on the field that
can alter the field angular momentum.
Indeed this has been exploited in a number of recent
experiments~\cite{TwistedPhotons}.
\par
\section{Summary}
\par
In a previous work, we considered a dielectric medium with an index of
refraction that was time-independent and spatially uniform, and the
dielectric was covered with a thin gradient-index antireflection
coating \cite{BICB}.
We found that the total energy and the Gordon momentum were conserved
quantities (constant in time). 
In this work we have relaxed these conditions to include a dielectric
medium that has a spatially varying index of refraction that is constant
in time.
The fact that the index of refraction is constant in time essentially
means that the dielectric medium is mechanically rigid and subject to
an external constraint.
This constraint means that we are not dealing with a closed system.
Using the continuity equations for energy density and momentum density,
we derived a symmetric energy-momentum tensor for the electromagnetic
field. Due to the fact that the system is not closed, the divergence of
the stress-energy tensor is not zero, but equal to a generalized
Helmholtz force vector that represents the constraint.
Similarly, the divergence of the angular momentum density is not zero,
due to the external constraint of a time independent index of
refraction.
We found that the total energy, Gordon momentum, and total angular
momentum are not conserved because of the external constraint on time
independence of the index of refraction.
However, the time dependence of the total energy, Gordon momentum, and
total angular momentum is related to the constraint, which is
proportional to the gradient of the index of refraction.
\par
Finally, we note that time $t$ has been renormalized
to ${\bar t} = t/n({\bf r})$ in the intermediate steps used to obtain
the symmetric energy-momentum tensor and its continuity relation given
by its four-divergence in Eq.~(\ref{EQq2.09}).
However, the final conservation laws of total energy, total momentum
and total angular momentum, given by Eqs.~(\ref{EQq3.04}),
(\ref{EQq3.03}), and (\ref{EQq3.09}), respectively, are
expressed in terms of the unrenormalized time $t$.


\begin{thebibliography}{}
\bibitem{RigidFootnote}The assumption of a rigid dielectric medium does not violate relativistic principles, however, it leads to a non-zero four divergence of the stress--energy tensor for the electromagnetic field, since the electromagnetic field is not a closed system (because it is interacting with the dielectric medium).
\end{thebibliography}
\end{document}